# Event Detection from Social Media Stream: Methods, Datasets and Opportunities

Quanzhi Li
*Tencent*
Bellevue, USA
quanzhili@global.tencent.com

Yang Chao
*Tencent*
Shenzhen, China
youngchao@tencent.com

Dong Li
*Tencent*
Shenzhen, China
saulli@tencent.com

Yao Lu
*Tencent*
Shenzhen, China
loyolu@tencent.com

Chi Zhang
*Shanghai World Foreign Language Academy*
Shanghai, China
zhangchi@stu.swfla.org

*Abstract*—Social media streams contain large and diverse amount of information, ranging from daily-life stories to the latest global and local events and news. Twitter, especially, allows a fast spread of events happening real time, and enables individuals and organizations to stay informed of the events happening now. Event detection from social media data poses different challenges from traditional text and is a research area that has attracted much attention in recent years. In this paper, we survey a wide range of event detection methods for Twitter data stream, helping readers understand the recent development in this area. We present the datasets available to the public. Furthermore, a few research opportunities are discussed as potential future research directions.

## I. INTRODUCTION

The rapid development of social media platforms has led to an explosion of user-generated data posted on the Internet. The huge amounts of such data have enabled the study of many research problems, and event detection is one of the important topics. Twitter is a fast communication channel for spreading breaking news and events, and a good resource for detecting real-time events, such as an earthquake, bombing, or strike event. In this paper, we survey the techniques found in the literature for event detection from Twitter data stream. We also provide the available datasets and discuss several research opportunities.

Event definitions vary slightly in previous studies. McMin et al. [66] defines event as "something significant that happens at specific time and place". Xie et al. [96] define events as "real-world occurrences that unfold time and space", which can be described with the so-called 4Ws (what, who, when & where). In this study, we use the definition from Allan et al. [4, 5] and Topic Detection and Tracking (TDT) project: events are real-world occurrences that unfold over space and time, and the objective of event detection is to discover new or previously unidentified events. This definition is more general and has been used in many studies [4, 5, 90, 96, 98, 8, 41, 42, 43, 44, 45, 87].

Event detection from conventional media has been long addressed in the TDT program. However, event detection from social media poses new challenges that are different from those in traditional media. In contrast with the well-written and structured news articles, tweets are restricted in length and the textual information is very limited. The messages include large amounts of informal and abbreviated words, spelling and grammatical errors, irregular sentence structures and mixed languages. Tweets also contain large amounts of meaningless messages, spams, advertisements, and rumors [16, 37, 38, 52, 53, 8, 41, 40, 48, 49], which will negatively affect the detection algorithm performance. Event detection from Twitter data streams involves techniques from various areas, such as natural language processing, text mining, information retrieval (IR) and social network analysis [51, 70].

In this paper, we do not provide an exhaustive review of existing methods but choose the representative techniques to give readers a perspective on the main research directions. The main differences between this paper and previous surveys [59, 8, 76] on event detection from social media data are:

1. Many datasets used in the event detection studies are not available to the public. Previous survey papers do not provide a review on which dataset is available to the public. We checked previous studies to see if their used datasets are available to the public and introduce all the datasets that are available.
2. We provide a full review of the evaluation metrics for event detection. None of the previous surveys has done this; they only mentioned a couple of them in their papers.
3. We also reviewed the recent studies utilizing neural network. Previous surveys have not reviewed the approaches based on neural network.
4. Based on the survey, our research experience and work experience with event detection applications, we provided and discussed a list of challenges and opportunities in this field, which could be potential future research topics.

Though we mainly surveyed the event detection techniques on Twitter data, we think the structure we classify the detection methods, the evaluation metrics and the discussions on future research opportunities are also applicable to the short messages produced on other social media platforms.

## II. Event Detection in Twitter Data Stream

We can review the event detection methods from different angles. Based on the event type, these methods can be classified into unspecified vs. specified event detection. In TDT, event detection can be broadly classified into two categories: new event detection (NED) and retrospective event detection (RED). NED is the discovery of new events from data streams in real time [4], and RED focuses on discovering previously undetected events from historical collections [98]. NED is also called a first-story detection or novelty detection [79, 80, 81, 41, 42].

Table 1 organizes the representative detection approaches based on the event type, detection task, detection technique and their application. Since most event detection techniques are for unspecified events and NED, and most unspecific event detection are for NED, therefore, in the following subsections, we focus more on unspecified event detection.

### A. Unspecified Event Detection

For unspecified events, we have no prior information about the events, so the unspecified event detection techniques rely mainly on exploiting the temporal patterns or signal from Twitter data streams. Unspecified events of interest are usually driven by breaking news, emerging events and general topics attracting the attention of a large number of users.

Fig. 1 shows the typical workflow for a NED system for unspecified events. It includes not only the detection part, but also the post-detection components, which are necessary for most event detection applications. Specified event detection workflow has the similar architecture.

1. **Preprocessing stage**. The noise filtering component is to remove spam tweets, and tweets that are basically nonsense, such as profanity, chitchat, and advertisement. Noise filter is usually built as a classifier [53, 41, 42, 87]. The metadata extraction component extracts entities or other metadata (e.g., geo-location, links, and hashtags) that might be used in later stages.
2. **Event detection stage**. Depending on the event detection type (specified, unspecified, RED, NED), the actual detection technique (e.g., clustering based, term based, retrieval based) in the detection stage may be different. The cluster defragmentation and cluster purging components are to merge relevant event clusters together and purge old events from memory; they may not be needed for some detection techniques [41, 42].
3. **Post-detection stage**. Depending on the application, the post-detection stage may need some components. For example, event summarization may be necessary for most use cases, and the newsworthiness ranking, and event veracity identification (rumor detection) components will benefit news agency users [41, 53, 40, 47, 48, 49, 50, 71, 52].

We organize the unspecified event detection methods into the following categories: clustering based, term based, and neural network based. The neural network approach has overlap with others; we use a separate type for it to highlight the recent studies exploiting neural networks.

#### 1) Clustering Based Approaches

Due to the unpredictability and dynamicity in social media data streams, there was a tendency in previous studies to use unsupervised methods, such as clustering and tensor decomposition for event detection [33]. Many event detection algorithms tackle the problem as a stream clustering task. Becker et al. [11] have used an incremental clustering algorithm to detect events from the Twitter stream. Petrovic et al. [2010] and Wurzer et al. [95] used a Locality Sensitive Hashing (LSH) to detect and cluster events from high-volume tweet streams in constant time and space. Aggarwal and Subbian [2] proposed a stream-based clustering algorithm on each incoming post. McCreadie et al. [64] showed that K-means clustering can be successfully used for event detection. Many other approaches have utilized hierarchical or incremental clustering approaches [22, 41, 69, 67]. Corney et al. [22] proposed clustering word n-grams, Li et al. [41] proposed clustering semantic terms, and Morabia et al. [67] proposed clustering segments. Nguyen et al. [69] clustered term frequency-inverse document frequency (tf-idf) vectors after identifying candidate clusters using entity inverted indices. Fedoryszak et al. [26] represent an event as a chain of clusters over time. Their algorithm design is based on the realization that they can decompose burst detection and clustering into separate components that can be scaled independently. Wang and Zhang [92] build a joint model to filter, cluster, and summarize the tweets for new events.

Online clustering-based approaches are prone to cluster fragmentation and are usually unable to distinguish between two similar events occurring around the same time [79, 66, 36].

#### 2) Term Based Approaches

Clustering based approach is a document-pivot technique because it relies on tweet, which is a short document. The term-based approach is a feature-pivot technique. It models an event in text streams as a bursty activity, with certain features rising sharply in frequency as the event emerges. The clustering based and term-based approaches can be used together.

TwitInfo [61] uses a streaming algorithm to detect spikes in tweet data, and the peak generated by the high volume of posts are considered as events. TwitterMonitor [63] detects emergent topics by identifying the bursty terms within a small-time window. If the system detected high frequency terms co-occur in many tweets in the given time window, they are placed in the same group. Similarly, enBlogue [6] computes statistical values for tag pairs within a given time window and monitors unusual shifts in the tag correlations to detect emergent topics. TopicSketch [97] detects bursty topics by relying on the concept of word acceleration. Some studies utilize anomaly detection algorithm, whose technique is like the term based bursty detection technique. Li and Zhang [45] exploit the semantic types of event related terms. An event is usually defined by the 4Ws questions: who, what, where and when. An event tweet usually contains terms corresponding to these aspects, and these terms can be classified into different semantic classes/types, such as entity names (who) and location (where). They also use the semantic terms for event summarization.

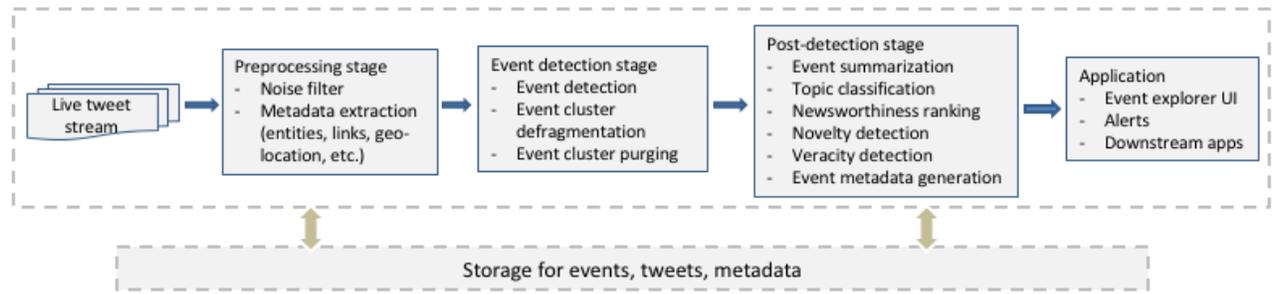

Fig. 1. The typical workflow for a NED system for unspecified events. * For RED, the data source will not be a live tweet stream.

TABLE I. REPRESENTATIVE EVENT DETECTION STUDIES. **S** - SPECIFIED, **U** - UNSPECIFIED, **NN** - NEURAL NETWORK, **LDA** - LATENT DIRICHLET ALLOCATION, **LSH** - LOCALITY SENSITIVE HASHING, **CRF** - CONDITIONAL RANDOM FIELDS, **SVM** - SUPPORT VECTOR MACHINE, **SVD** - SINGULAR VALUE DECOMPOSITION

| Paper | Event type | | Detection task | | Technique | Application |
|---|---|---|---|---|---|---|
| | S | U | NED | RED | | |
| Petrović et al., 2010 [79] | | x | x | | online clustering & LSH | general event detection |
| Long et al., 2011 [58] | | x | x | | hierarchical clustering | general event detection |
| Massoudi et al., 2011 [62] | x | | x | x | generative language modeling | query-based event retrieval |
| Metzler et al., 2012 [65] | x | | x | x | temporal query expansion | query-based structured retrieval |
| Weng and Lee 2011 [94] | | x | x | | wavelet analysis & graph | general event detection |
| Becker et al., 2011a [11] | | x | x | | online clustering & SVM | general event detection |
| Cordeiro, 2012 [21] | | x | x | | wavelet analysis & LDA | general event detection |
| Popescu & Pennacchiotti, 2010 [83] | x | | x | | decision tree | controversial events |
| Phuvipadawat & Murata, 2010 [82] | | x | x | | online clustering | breaking news detection |
| Benson et al., 2011 [14] | x | | | x | factor graph model & CRF | musical event detection |
| Lee & Sumiya, 2010 [38] | x | | x | | modeling of crowd behavior | geosocial event monitoring |
| Sakaki et al., 2010 [86] | x | | x | | SVM | disaster events monitoring |
| Becker et al., 2011b [11] | x | | x | x | recursive query construction | query-based event retrieval |
| McCreadie et al., 2013 [64] | | x | x | | k-mean clustering & LSH | general event detection |
| Li et al., 2012 [39] | | x | x | | user diversity-based measures | general event detection |
| Abdelhaq et al., 2013 [1] | | x | x | | keywords & spatial clustering | general event detection |
| Parikh & Karlapalem, 2013 [78] | | x | x | | hierarchical clustering | general event detection |
| Corney et al., 2014 [22] | x | | x | | online clustering | sports |
| Chen et al., 2014 [19] | | x | x | | temporal topic modeling | general event detection |
| Guille & Favre, 2015 [28] | | x | x | | mention, anomaly detection | general event detection |
| Nur'Aini et al., 2015 [73] | | x | x | | k-mean clustering & SVD | general event detection |
| Zhang & Qu, 2015 [99] | | x | x | | hierarchical clustering | general event detection |
| Wang et al., 2015 [91] | | x | x | | online clustering, summarization | general event detection |
| Wurzer et al., 2015 [95] | | x | x | | clustering & LSH | general event detection |
| Hasan et al., 2016 [31] | | x | x | | clustering & LSH | general event detection |
| Xie et al., 2016 [97] | | x | x | | SVD, clustering | general event detection |
| Stilo & Velardi, 2016 [88] | x | | x | | hierarchical clustering | general event detection |
| Li et al., 2017 [41] | | x | x | | NN, semantic terms clustering | general event detection |
| Chen et al., 2017 [18] | | x | x | | NN, online clustering | general event detection |
| Wang & Zhang, 2017 [92] | | x | x | | NN, multitask learning, clustering | general event detection |
| Edouard, 2018 [24] | x | | x | | domain vocabulary and KB | soccer domain |
| Comito et al., 2019 [17] | | x | x | | text & temporal feature clustering | general event detection |
| Fedoryszak et al., 2019 [26] | | x | x | | cluster chain, graph | general event detection |
| Nguyen et al., 2019 [69] | | x | x | | entity clustering | general event detection |
| Morabia et al, 2019 [67] | | x | x | | segments hierarchical clustering | general event detection |
| Saeed et al., 2019 [85] | | x | x | | graph, clustering | general event detection |
| Hettiarachchi et al., 2020 [33] | | x | x | | NN, hierarchical clustering | general event detection |
| Cao et al., 2021 [15] | | x | x | | GNN, clustering | general event detection |
| Li and Zhang, 2021 [45] | | x | x | | semantic term, GCN, clustering | general event detection |

Term based can often capture misleading term correlations and measuring the term correlations can be computationally prohibitive in an online setting approach [63, 88, 78, 45].

*3) Approaches Using Neural Network*

Recent studies have applied neural network and deep learning technologies on event detection from social media [17, 18, 92, 33, 41, 42, 68, 33, 15, 43, 45].

The embeddings, such as word embedding or tweet embedding, learned from text via neural network technologies capture the semantic and syntactic regularities in the text. It solves the vocabulary mismatch problem existing in the traditional event detection approaches. Deep network also helps us learn the latent information embedded in text. Chen et al. [18] proposed a deep neural network-based approach where tweets were converted into fixed length vectors using pretrained GloVe embeddings [77], which is then used for tweet clustering. Wang and Zhang [92] build a joint model to filter, cluster, and summarize the tweets for new events. Tweet representation built from Long Short-Term Memory is shared among filtering, clustering, and summarization. Hettiarachchi et al. [33] propose a novel method termed Embed2Detect by combining the characteristics in word embeddings and hierarchical clustering. We expect more studies utilizing neural network will appear. Car et al. [15] propose a novel Knowledge-Preserving Incremental Heterogeneous Graph Neural Network (KPGNN) for incremental social event detection. To acquire more knowledge, KPGNN models complex social messages into unified social graphs to facilitate data utilization. To continuously adapt to the incoming data, KPGNN adopts contrastive loss terms that cope with a changing number of event classes. To deal with large social streams, KPGNN periodically removes obsolete data to maintain a dynamic embedding space.

The drawback with deep neural network is that it may have velocity issue when used for online NED if the network structure is very complex.

*B. Specified Event Detection*

Specified event detection includes known or planned social events, or events related to some specific topics. These events could be partially or fully specified using related content or metadata information, such as location, venue, or keywords. The detection task could be NED or RED. The techniques used can be IR-based or the ones described in previous section, such as the term based approach. Because the events are specified and it is easier to build training data than unspecified events, many specified event detection approaches use supervised detection algorithms [86, 83, 84, 14, 22, 65, 24]. For example, Sakaki et al. [86] formulated event detection as a classification problem and trained an SVM classifier to detect earthquakes and typhoons events.

*C. NED and RED*

Depending on the task and the type of event, event detection in Twitter can also be classified into RED and NED. Because NED techniques involve continuous monitoring of Twitter data for discovering new events in near real time, they are naturally suited for detecting unknown real-world events and breaking news [79, 80, 11, 95, 41, 33, 45, 54]. NED techniques can also be used for specified event detection, although most studies focus on unspecified events. When the task involves specific events (e.g., disasters, crimes, sports) or a specific information about the event (e.g., specific organization, person, or location), this information could be integrated into the NED methods by using classification or filtering techniques [86, 83, 37, 65, 22, 88].

While most research focused on NED to exploit the timely information provided by social data streams, there are also interests in RED from historical data [62, 65, 14, 11]. For RED, because in most cases we already have the whole data collection, traditional IR-based methods can be exploited. Most methods for NED can also be utilized for RED with just small changes [8]. Twitter provides limited search capabilities that allow to retrieve tweets, so some RED tasks are conducted by searching old tweets from Twitter. Vocabulary mismatch is a problem in this case, since Twitter does not provide embedding search.

III. DATASETS

Evaluation datasets are important for comparing and evaluating the effectiveness of different event detection approaches. One issue with the event detection from Twitter stream is that many datasets used by the studies are not available to the public, and different studies used different datasets. Therefore, it is hard to say which approaches have the state-of-the-art performance. We collected a list of Twitter datasets that are available to the public. Below are the brief introduction and link of the datasets that are available to the public.

- Dataset 1: This dataset covers three topics: FA Cup Final, Super Tuesday for US Elections, and US Elections [3]. They have 13, 8 and 26 events, respectively. The tweets were collected from Nov 2012. *http://socialsensor.iti.gr/results/datasets*

- Dataset 2: This one consists of 41 events and 671K tweets posted within the area of Manhattan, NYC during 12/2014 [17]. The events are about general topics. *https://dl.acm.org/doi/10.1145/3332185*

- Dataset 3: this one contains two events: MUNLIV – events about English Premier League on October 20, 2019 between Manchester United and Liverpool, And BrexitVote – events about Brexit Super Saturday 2019 on October 19, 2019 [33]. *https://github.com/hhansi/twitter-event-data-2019*

- Dataset 4: this one has two sub datasets, one from the earthquake domain and another from DDoS attack domain [92]. The tweets were collected from June 2013 to April 2016. https://github.com/wangzq870305/joint_event_detection

- Dataset 5: this one consists of 27 topics and 116K tweets from April till September 2011 [80, 81, 95]. *https://era.ed.ac.uk/handle/1842/7612*

- Dataset 6: Inouye and Kalita [34] collected the top trending topics from Twitter for the year of 2011, and finally got 50 trending topics with a total set of 75K tweets. *https://ieeexplore.ieee.org/document/6113128*

- Dataset 7: this corpus is provided by McMinn et al. [66]. The tweets were collected from Dec. 2012. It has 506 events on different general topics containing over 150K relevant tweets. The problem with this dataset is that it contains only tweet id, and the majority of these tweets cannot be downloaded from Twitter since they are not available any more. *http://mir.dcs.gla.ac.uk/resources/*

## IV. EVALUATION METRICS

To evaluate the quality of the detected events, various metrics have been used in previous studies, summarized below.

### A. Normalized Topic Weighted Minimum Cost (Cmin)

This metric is from TDT program [5] and has been used by several studies [80, 81, 95, 92]. *Cmin* is a linear combination of miss and false alarm probabilities, which allows comparing different methods based on a single value metric. Computing *Cmin* needs several equations, and we skip them here due to the space limit. See [5] for more details.

The TDT project assumed that the documents come from a noiseless stream, such as newswire, which means that all the documents in the stream are considered newsworthy. As a result, evaluation based on *Cmin* has ignored precision and focused instead only on miss and false alarm rate. However, social media stream is very noisy, which means that *Cmin* is no longer a good metric here. To get a complete picture of the effectiveness of an event detection approach, we should measure both recall and precision, described below.

### B. Precision, Recall and F-measure

These three metrics could be used if a labeled dataset is used to evaluate the performance of an algorithm. An event recall is the percentage of ground-truth events successfully detected by a method. A ground-truth event is considered successfully detected if there exists a predicted event that matches certain number of tweets or terms (threshold varies by studies). Precision is defined as the percentage of ground-truth events in the generated events. F measure is the harmonic mean of precision and recall. Many previous studies [3, 17, 80, 32, 85] have used part or all these three metrics.

The issue with event cluster level precision, recall and F is that they cannot measure the cohesiveness within a cluster. To overcome this drawback, we suggest using the following two measures: NMI and B-Cubed.

### C. Normalized Mutual Information (NMI)

NMI [60, 89] and C-Cubed [Amigo et al., 2008] have been used in previous studies on general and social media message clustering [12, 93, 23, 26, 41, 42]. We chose them because both metrics balance the desired clustering properties: to maximize the homogeneity of events within each cluster, and to minimize the number of clusters that tweets of each event spread across.

NMI is an information-theoretic metric that was originally proposed as the objective function for cluster ensembles. It measures how much information is shared between actual ground truth events, each with an associated tweet set and the clustering assignment. More details are in [89].

### D. B-Cubed

B-Cubed [9] estimates the precision and recall associated with each tweet in the dataset individually, and then uses the average precision $P_b$ and average recall $R_b$ values for the dataset to compute B-Cubed:

$$B\text{-}Cubed = \frac{2 * P_b * R_b}{P_b + R_b} \quad (1)$$

For each tweet, precision is defined as the proportion of items in the tweet's cluster corresponding to the same event, and recall is the proportion of tweets that correspond to the same event, which are also in the tweet's cluster.

## V. CHALLENGES AND OPPORTUNITIES

Social media post is made up of short, noisy, and unstructured data, and the volume is huge. These challenges have been well discussed in previous studies. Based on the review of related studies, our research experience and our direct work experience with real world applications and the stakeholders, such as news agency, public safety office and big corporation, we present the following challenges and opportunities in this field, which could be potential future research topics. Due to the space limitation, we just briefly discuss them.

### A. Event Evolution Stages

An event may evolve and develop into multi stages. For example, a bombing attack event may have a few stages: the booming incident, pursuit of the suspect, the arrest of the suspect, and sentence of the suspect. Depending on the application and the preferred granularity level, these stages may also be considered as different but related events. As the event evolves, the terms used to describe the event may also gradually change. Take the tsunami in Japan that occurred in 2011 as an example, initially, the event is dominated by keywords like "earthquake" and "tsunami", but later words such as "nuclear" and "radiation" are introduced. Clearly identifying the development stages of an event will help us analyze, understand, present, and organize the event. There are some explorations on this topic [75, 26], but given the importance and challenge of this problem, how to identify these evolution stages and connecting them together has not attracted enough research intention.

### B. Multi-task Learning

Studies already show that jointly learning can improve performance of tasks that are related or share some common information [92, 44, 48, 56]. In the event detection workflow, depending on the applications, the following tasks might be involved: event detection, entity extraction, event summarization, topic classification, rumor detection, and novelty detection. One future research direction is to explore multi-task learning techniques on these tasks. These tasks share some information and some of them also have inter-dependence relation. We expect jointly learning will benefit at least some of them, as initially demonstrated by [92]. The recent advances in neural network and deep learning technologies will also help this exploration.

## C. Temporal Information Identification

In social media, users may talk about any event; some events may be as old as days, months, or even years ago, e.g., a discussion about an event occurring in World War II. A real-time novel event detection system is only interested in events that are happening now or just happened a short time ago. To filter out the old events, we need to identify the temporal information in a cluster's tweets and use that information to determine whether the event is a new one. When we say an event is an "old" event, it may have different meanings in different use cases or applications. Many events need to specifically extract the temporal information from its tweets to decide if it is an old event or not; the traditional novelty detection techniques may not work for this case. Li et al. [2017] and Li and Zhang [2021] identify temporal information and use it as one semantic type in their clustering algorithms. But they did not explicitly address the issue mentioned above.

## D. Event Witness Identification

Social media has provided citizen journalism with an unprecedented scale, and access to a real time platform, where once passive witnesses can become active and share their eyewitness testimony with the world, including with journalists who may choose to publicize their report. Identifying witness accounts is important for rumor debunking, crises management, and basically any task that involves on the ground eyes. Witness identification involving analyzing the tweet text, location information in user profile, messages posted preceding and after the message of this event. Fang et al. [25] use n-grams and traditional classifiers to identify witness accounts of an event, the proposed method was used in debunking rumors in social media [53, 41, 42]. This area is under-researched, and methods exploiting neural networks may benefit.

## E. Multimodal Event Detection

Currently most event detection studies in Twitter focus on text content, but more and more social media messages contain image, video, voice, or links. As advances in video, voice and image analysis, multi-modality algorithms have been utilized in other applications and shown success [53, 47]. One promising research direction is to exploit the multimedia information in event detection on Twitter. Reference [10] and [100] have explored this direction, but only in a narrow domain. Like the case of multi-task learning, the recent advances in neural network technologies will help this research direction.

## F. Event Popularity Prediction

Many events develop gradually, unnoticed at its early stage and finally evolve to an event having big impact. Detecting an event when it is already spreading or going viral is not hard. One challenging and important task is to predict the event's popularity, so that the related parties can get alerts earlier and get prepared or act before it causes series damage. One issue with most current detection approaches is that when an event has not evolved for some time, that event may be removed from the radar of the system, usually due to the computing resource constraint, but later it becomes a big event. Including the popularity prediction ability in the whole event detection workflow will help. Popularity prediction involves not just the textual information, but also network propagation and social media user profile information. Gupta et al., [29] uses regression classification with social and event features to predict even popularity, while Chen et al. [20] use just hashtags. One interesting direction would be to explore both neural network models and a large set of multimodal features.

## G. Rumor Detection and Event Detection Integration

Rumor early detection is to detect a rumor at its early stage before it wide spreads on social media, so that one can take appropriate actions earlier. Early detection is especially important for a real-time system, since the more a rumor spreads, the more damages it causes [40, 41, 47, 48, 54, 52]. Currently, rumor detection and event detection are two separate tasks. After an event is detected, then we detect the veracity of that event. One challenging research direction is to detect the event and rumor jointly, so we can identify the veracity of the event as early as possible. Much information can be shared by these two tasks, such as the entities extracted, user info and network propagation info. We think this will be an interesting and challenging research topic, and a good solution will have very big impact on the rumor detection field.

## H. Cross-platform and Cross-language

Most previous studies on event detection on social media focus on only one specific social media platform. A solution that can detect and link events on different platforms will provide us at least two benefits: 1. The same event may have different burst or propagation velocity and characteristics on different platforms. The knowledge about this event gained from one platform may help us detect and analyze the event on another platform. 2. For the same event, user responses and opinions may be different on different platforms. A cross-platform solution may help related parties to gain a deep understanding and full picture about people's responses to an event, such as a public safety event. Cross-language event detection and analysis has becoming much more attractive in recent years, since nowadays more events have world-wide effect, such as events about finance, politics, and public health crisis. Like the cross-platform event detection case, a cross-language solution will also benefit both the detection and understanding of an event. Liu et al. [57] unify multi-lingual sources into same language, and then detect events by merging the same entities and similar phrases and present multiple similarity measures by using word2vec model. For multi-lingual event detection, this study translates different languages into one, and it do not do anything more than that.

## VI. CONCLUSION

In this paper, we survey a wide range of event detection methods for Twitter data stream and present a list of datasets that are available to the public. A few research opportunities are also discussed, which could be potential future research directions.